\documentstyle[a4,12pt,psfig]{article}
\newcommand{\al}{\alpha}
\newcommand{\bt}{\beta}

\newcommand{\be}{\begin{equation}}
\newcommand{\ee}{\end{equation}}
\newcommand{\ba}{\begin{eqnarray}}
\newcommand{\ea}{\end{eqnarray}}
\newcommand{\de}{\delta}
\newcommand{\dd}{\partial}
\newcommand{\ga}{\gamma}
\newcommand{\sg}{\sigma}

\newcommand{\nn}{\nonumber}

\newcommand{\vsa}{\vspace{0.4cm}}
\newcommand{\vsb}{\vspace{0.8cm}}

\newcommand{\eps}{\varepsilon}
\newcommand{\vfi}{\varphi}

\newcommand{\pijl}{\leftrightarrow}
\begin{document}
\begin{titlepage}
\begin{flushright}
hep-th/9606011 \\
THU-96/22 \\
May 1996
\end{flushright}
\vsa
\begin{center}
{\large\bf The Torus Universe in the Polygon Approach to 2+1-Dimensional
Gravity\vsa\vsb\\}
           M. Welling\footnote{E-mail: welling@fys.ruu.nl}\\
   {\it Institute for Theoretical Physics\\
     University of Utrecht\\
     Princetonplein 5\\
     P.O.\ Box 80006\\
     3508 TA Utrecht\\
     The Netherlands}\vsb\vsa\\
\end{center}
\begin{abstract}
\noindent
In this paper we describe the matter-free toroidal spacetime in 't Hooft's
polygon approach to 2+1-dimensional gravity (i.e. we consider the case without
any particles present). Contrary to earlier results in the literature we find
that it is not possible to describe the torus by just one polygon but we need
at least two polygons. We also show that the constraint algebra of the polygons
closes.
 \end{abstract}

\end{titlepage}
\section*{Introduction}
As is well known nowadays, gravity in 2+1 dimensions is flat everywhere outside
sources
\cite{eerst, beginarticle}. This means that the gravitational field itself has
no local degrees of freedom. One can make the theory non-trivial by adding
sources (e.g. point-particles) or considering a non-trivial topology of a
closed universe.\footnote{The topology of the universe under consideration is:
\[ {\cal M}=\Sigma(g)\times R\]
where $\Sigma(g)$ is a genus-g spacelike surface and $R$ is in the time
direction.}
For N point-particles that live on a genus-g surface for instance, the phase
space is $12g-12+4N$ dimensional \cite{tHooft2, Witten, Carlip1}. This formula
is however wrong in the case of a torus in the absence of particle sources.
This is due to the fact that the torus has some symmetries because of which the
counting argument breaks down. The toroidal universe, with or without
cosmological constant, has been extensively studied in the past. Its classical
solutions \cite{Moncrief, Nakao} as well as the quantum theory \cite{Carlip1,
Carlip2, Louko, Waelbroeck1, Nelson, Ezawa} are well understood in both the ADM
formalism and in the Chern-Simons formalism. From this work we know that the
dimension of the space of ADM solutions is four (i.e. there are only four
independent degrees of freedom) The torus is therefore a particularly simple
model and a convenient starting point for a quantization program. As we are
interested in the polygon description of 2+1-D gravity invented by 't Hooft
\cite{tHooft1, tHooft2} we decided to study the torus in this approach.
Guadagnini and Franzosi  had already worked on this problem \cite{Franzosi}.
But their counting of degrees of freedom was a bit puzzling to us. We found
that they described a subset of all possible solutions for a torus universe.
This is due to the fact that they use only one polygon for their slicing of
spacetime. This is not enough to cover all possible tori. The simple solution
to this problem is to add another polygon. This unfortunately implies that the
description loses its simplicity due to the fact that polygon transitions may
take place during evolution. This fact also considerably complicates the
quantization. The temporary conclusion is that the polygon approach is not the
most convenient description for the matter-free torus universe as compared with
other approaches.\\

\noindent
In section 1 we recapitulate the way Carlip describes a toroidal spacetime and
stress the fact that the phase space is 4 dimensional.\\
Section 2 contains an introduction to the polygon approach. We compute the
constraint algebra of the polygons and conclude that the algebra closes but is
highly nonlinear. We propose to define a new constraint for which the
constraint algebra closes linearly.
In this section we also reproduce the one-polygon solution for the toroidal
universe of
Guadagnini and Franzosi and it is shown that it contains only part of phase
space.\\
In section 3 we propose a 2 polygon representation for the torus and show that
the phase space is now 4 dimensional.\\
In the discussion we comment on possible roads to quantization.\\
Appendix A gives some details of the calculation of the constraint algebra.\\
In appendix B we carefully count the number of degrees of freedom for the
two-polygon torus.

 \section{ The Torus Universe}
In this section we recapitulate the work of Carlip \cite{Carlip1} and Louko and
Marolf \cite{Louko}. The construction of the torus starts by studying its first
homotopy group (or fundamental group):
\be
  \pi_1(T^2\times R)=\pi_1(T^2)=Z\oplus Z
\ee
The fact that the fundamental group is isomorphic to $Z\oplus Z$ implies that
there are 2 closed paths $\al_1$ and $\al_2$ that cannot be deformed into one
another or in the trivial path. They fulfill the relation:
\be
\al_1\circ \al_2\circ \al_1^{-1}\circ \al_2^{-1}=I
\ee

\begin{figure}[t]
\centerline{\psfig{figure=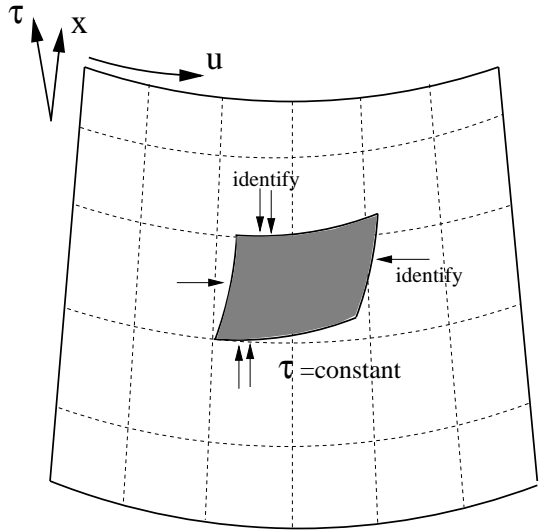,angle=-90,height=7cm}}
\caption{The torus as constructed by Carlip.}
\label{torus1}
\end{figure}

\noindent
where $\circ$ stands for path composition. Next we must construct the
representation of this fundamental group in the three dimensional Poincar\'e
group ISO(2,1). From Witten's gauge formulation of 2+1-D gravity \cite{Witten}
we know that 2 holonomies around a loop  are equivalent if they can be
conjugated into each other (i.e. $\Lambda\sim S\Lambda S^{-1}$). Louko and
Marolf have shown that the possible holonomies (up to conjugation) split into 4
sectors; a spacelike sector, a timelike sector, a null sector and a static
(spacelike) sector. We will not consider the timelike and null sector in the
following as they suffer from causality problems (closed timelike curves).
The spacelike sector can always be conjugated into:
\ba
\Lambda(\al_1)&=&B_y(\eta_1)T_x(a_1)\label{transdyn}\\
\Lambda(\al_2)&=&B_y(\eta_2)T_x(a_2)\nn
\ea
where $B_y(\eta_i)$ are boost matrices in the y-direction with rapidity
$\eta_i$, and $T_x(a_i)$ are translations in the x-direction over a distance
$a_i$.
The transformations $\Lambda(\al_i)$ are in  the spacelike sector because the
projection to SO(2,1) (i.e. $B_y(\eta_i)$) stabilizes a spacelike vector. \\
The static sector is given by:
\ba
\Lambda(\bt_1)&=&T_x(b_1)T_x(b_2)\nn\\
\Lambda( \bt_2)&=&T_x(b_3)\nn
\ea
Only these physical sectors are expected to be contained in the set of
``ADM-solutions''
\cite{Moncrief, Nakao}. The construction of the torus proceeds by finding a
non-degenerate triad $e^a_{~~\al}$ that reproduces these holonomies. Another
way is to find a fundamental region of Minkowski  space ${\cal J}$ upon which
the holonomies act properly discontinuously. The quotient space ${\cal
J}/\{\Lambda(\al_i)\}$ will then be our torus.\\
In the spacelike case this construction is as follows: first introduce the
coordinates:\footnote{Our definition of time is different from the one used in
Carlip's paper \cite{Carlip1}. We use $\tau$ where he uses $\frac{1}{\tau}$.}
\be t=\tau\cosh u~~~~~y=\tau\sinh u~~~~~x=x\ee
The Minkowski metric in these coordinates is:
\be
ds^2=-d\tau^2+\tau^2 du^2+dx^2
\ee
The transformations $\Lambda(\al_i)$ act on these coordinates as:
\be
\Lambda(\al_i): (\tau, u, x)\rightarrow (\tau, u+\eta_i, x+a_i)
\ee
The torus is thus constructed by identifying these points on a $\tau=$constant
surface (see figure (\ref{torus1})).

\begin{figure}[t]
\centerline{\psfig{figure=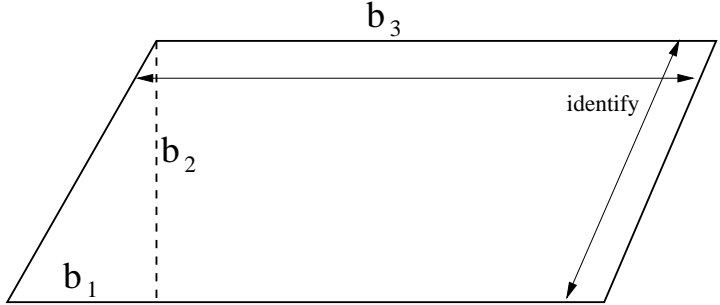,angle=-90,height=4cm}}
\caption{The static torus.}
\label{torus2}
\end{figure}

\noindent
In the case of the static torus  the holonomies act on the Minkowski
coordinates as:
\ba
&\Lambda(\bt_1):&(t,x,y)\rightarrow (t, x+b_1, y+b_2)\\
&\Lambda(\bt_2):&(t,x,y)\rightarrow (t,x+b_3, y)\nn
\label{transstatic}
\ea
In this case the torus is constructed by making these identifications on a flat
2-dimensional $t=$constant slice (see figure (\ref{torus2})). It is important
to notice that in the spacelike sector  it is possible that one of the $\eta_i$
is zero. Actually the construction works only if $\eps^{\al\bt}\eta_\al
a_\bt\neq 0$. This implies that the static torus is not in the set considered
by Carlip. As we will see, the static torus is included in the polygon
construction.

\section{ The Torus in the Polygon Approach}
\begin{figure}[t]
\centerline{\psfig{figure=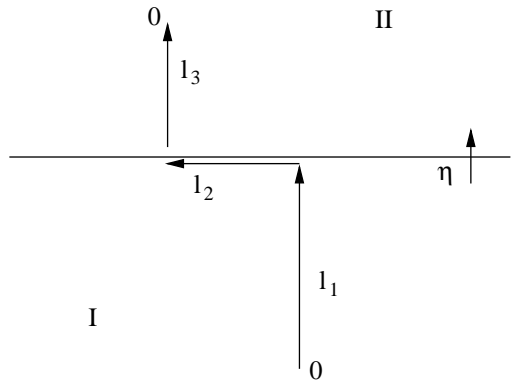,angle=-90,height=5cm}}
\caption{The The Poincar\'e transformation between frame I and frame II.}
\label{trans}
\end{figure}

In this section we give an introduction to the polygon approach and calculate
the constraint algebra of the polygons. We also construct a one-polygon torus
\cite{Franzosi}.\\
The polygon approach to 2+1-D gravity was invented by 't Hooft \cite{tHooft1,
tHooft2} as a Hamiltonian description of 2+1-D gravity.  The main idea behind
this approach is to construct a Cauchy surface by gluing together piecewise
flat patches of spacetime. Let's define coordinates $x^\mu_I$ on polygon I.
These can be transformed to a new frame (to be used in polygon II) by the
following Poincar\'e transformation (see figure (\ref{trans})):
\be
x^\mu_{II}=[T_y(-l_3)B_y(2\eta)T_x(-l_2)T_y(-l_1)]^\mu_{~~\nu}x^\nu_I
\label{Cauchy}
\ee
\noindent
where $B$ is a boost and $T$ is a translation. To construct a Cauchy surface we
consider the condition $t_I=t_{II}$. If we put this in (\ref{Cauchy}) we find
an equation for the boundary between the 2 frames:
\ba
y_s^I(t)&=&-vt+l_1~~~~~~~~v=\tanh\eta\\
y_s^{II}(t)&=&vt-l_3\nn
\ea

\begin{figure}[t]
\centerline{\psfig{figure=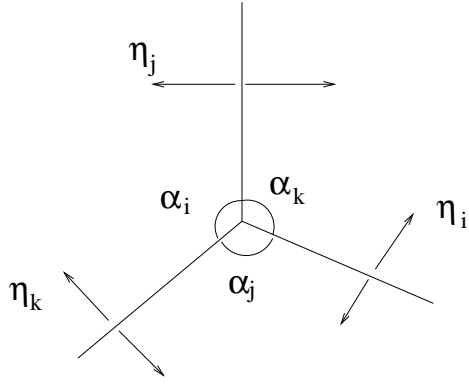,angle=-90,height=5cm}}
\caption{The three-vertex.}
\label{vertex}
\end{figure}
\begin{figure}[b]
\centerline{\psfig{figure=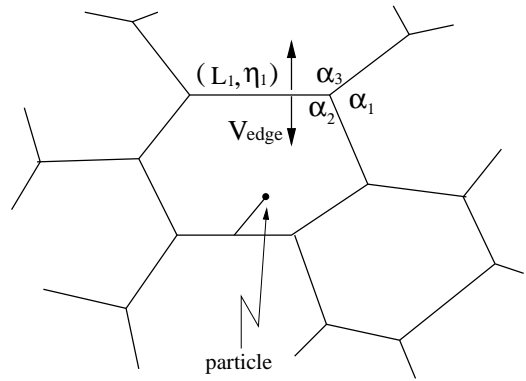,angle=-90,height=5cm}}
\caption{Cauchy surface made out of polygons.}
\label{pol}
\end{figure}

\noindent
where $y_s$ denotes the y-position of the boundary.
Next we define coordinates $x^\mu_{III}$ on a third frame and repeat the same
procedure. The boundary between II and III can now be calculated. The three
edges meet in a vertex (see figure (\ref{vertex})).
The fact that the three dimensional curvature must vanish at the vertex implies
a relation between the angles $\al_i$ and $\eta_j$:\footnote{The $\eta_j$
should however always obey the triangle relation:
$|\eta_i|+|\eta_j|\geq|\eta_k|~~~i,j,k=1,2,3$.}
 If we perform the three Lorentz transformations in sequel we should end up in
the original frame:
\be
x^\mu=[R(\al_k)B(2\eta_i)R(\al_j)B(2\eta_k)R(\al_i)B(2\eta_j)]^\mu_{~~\nu}x^\nu
\label{vertexrel}
\ee
The equations generated from this constraint are the vertex relations:
\ba
&& s_i:s_j:s_k=\sg_i:\sg_j:\sg_k\label{vertexrelations}\\
&& \ga_j s_k+s_i c_j+c_i s_j \ga_k=0\nn\\
&& c_i=c_j c_k-\ga_i s_j s_k\nn\\
&& \ga_i=\ga_j\ga_k+c_i \sg_j\sg_k\nn\\
&& \frac{c_j}{s_j}=-\frac{c_i}{s_i}\ga_k-\frac{\ga_j\sg_k}{s_i\sg_j}
\ea
where we defined:
\ba
\sigma_i&=&\sinh(2\eta_i)\label{def}\\
\gamma_i&=&\cosh(2\eta_i)\nn\\
c_i&=&\cos\al_i\nn\\
s_i&=&\sin\al_i\nn\\
\ea
They allow us to calculate for instance the three angles $\al_i$ from the three
rapidities $\eta_j$. If we continue to introduce new frames (and thus new
edges) we end up with a $t=$constant surface made out of polygons. (see figure
(\ref{pol})).

\begin{figure}[t]
\centerline{\psfig{figure=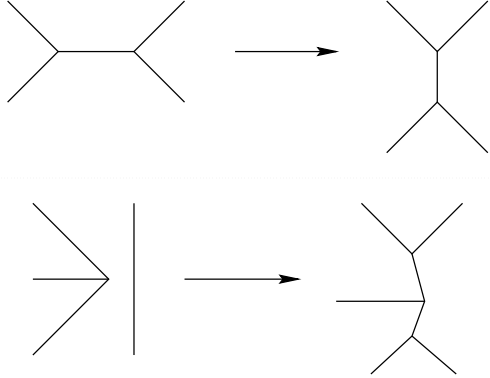,angle=-90,height=5cm}}
\caption{2 possible transitions.}
\label{transitions}
\end{figure}

\begin{figure}[b]
\centerline{\psfig{figure=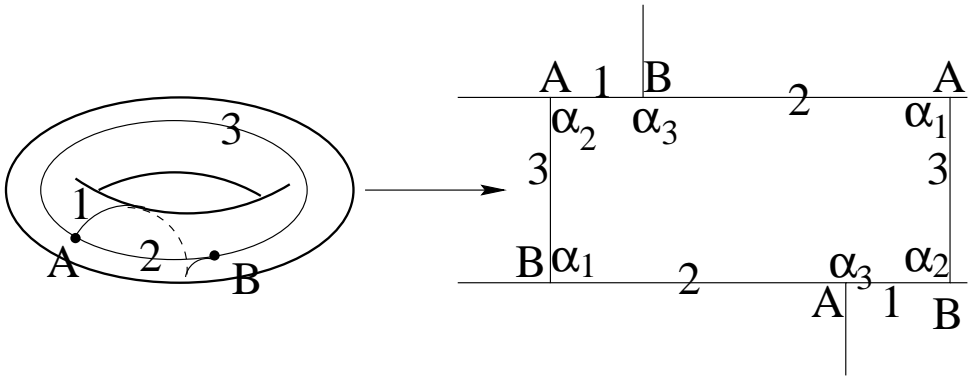,angle=-90,height=5cm}}
\caption{Cutting open a torus using only one polygon.}
\label{poltorus1}
\end{figure}

\begin{figure}[t]
\centerline{\psfig{figure=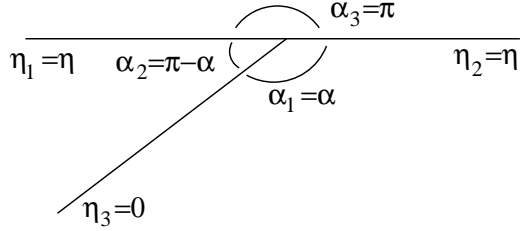,angle=-90,height=3cm}}
\caption{Vertex with two equal rapidities and one vanishing rapidity.}
\label{vertices}
\end{figure}

\noindent
Particles can also be included by putting them at the end of a boundary. The
transformation over that line is however a bit more complicated:
\be
\tilde{x}^\mu=[T(\vec{a})BRB^{-1}T(-\vec{a})]^\mu_{~~\nu}x^\nu
\ee
where $T(\vec{a})$ is a translation to the position of the particle $\vec{a}$,
$B$ is a boost in the direction of the velocity of the particle and $R$ is a
rotation over an angle proportional to the particle's mass. From the above we
may also notice that the coordinates are multivalued in the presence of the
particles (see \cite{Welling}). We will not go into this further as we will not
need particles in the following. The interested reader is referred to
\cite{tHooft1, tHooft2, Welling}. As was shown by 't Hooft we can turn this
description into a Hamiltonian formulation. The edge lengths $L_i$ are the
canonically conjugate variables to the rapidities $\eta_j$:
\be
\{2\eta_j,L_i\}=\de_{ij}
\ee
The Hamiltonian is then given by the sum of all the deficit angles at the
particles and the vertices:
\be
H=\sum_{P}\bt_P+\sum_V(2\pi-\al^V_1-\al^V_2-\al^V_3)
\ee
Remember that the angles $\al_i$ can be expressed in terms of the rapidities
$\eta_j$ using the vertex relations. The same is true for the angles $\bt_P$.
In the above expression $V$ labels the vertices and $P$ the particles. One can
now for instance check that the following relation hold:
\be
\frac{d}{dt}L_i=\{H,L_i\}~~~~~~~~~\frac{d}{dt}\eta_j=\{H,\eta_j\}~~~~~(=0)
\ee
Another important aspect is the constraints of the theory. First of all the
angles within one polygon should fulfill the relation:
\be
C_1=\sum_{k=1}^{N}\al_k-(N-2)\pi=0
\ee
where $N$ is the total number of corners inside the polygon. Also, if
temporarily we view the edge lengths $L_i$ as vectors, the sum of the edges
that enclose a polygon should add up to zero. In complex notation:
\be
C_2=\sum_{k=1}^{N}L_ke^{-i\theta_k}=0~~~~~~~~~~~\theta_k=
\sum_{l=2}^{k}\al_l-(k-1)\pi
\label{C2}
\ee
These are first class constraints and they generate ``gauge-transformations''
of the system. As can be verified $C_1$ pushes the particular polygon forward
in time. After this time evolution of one polygon, the surface must still be a
Cauchy surface. This can only be the case in general if the boundaries
rearrange themselves. So the phase space variables change as follows:
\be
\de L_i=\{C_1, L_i\}~~~~~~~~~~~~~~~~~~~\de \eta_j=\{C_1,\eta_j\}
\label{gauge}
\ee
The complex constraint $C_2$ generates boost transformations of the polygon (in
2 independent directions). If we boost the polygon to a new Lorentz frame, the
$L_i$ and $\eta_j$ change in order to keep the surface a $t$=constant surface.
This change is generated in the same way as in (\ref{gauge}).
For consistency we must check that the constraint algebra closes. The result of
this rather lengthy calculation is:
\ba
\{C_1^A,C_1^B\}&=&0\label{con-alg}\\
\{C_1^A,C_2^B\}&=&\frac{1}{2}(\frac{\sigma_1\gamma_N+\sigma_N c_1\gamma_1}
{s_1\sigma_N\sigma_1}+i\frac{\gamma_1}{\sigma_1})(1-e^{-iC_1^A})\de_{A,B}\nn\\
\{\bar{C}_2^A,C_2^B\}&=&\frac{1}{2}\frac{\gamma_1}{\sigma_1}
(C_2^A+\bar{C}_2^A)\de_{A,B}\nn
\ea
The index $A~~(B)$ labels the different polygons.
For the definition of the $\eta_i$, $L_i$ and $\al_i$ see figure
(\ref{definition}) in the appendix A where one can also find some details of
the calculation.
Although this algebra closes on-shell the first bracket is very nonlinear. To
remedy this situation we change $C_1$ to a new constraint:
\be
D_1=e^{iC_1}-1
\ee
The first bracket is now replaced by:
\be
\{D_1^A,C_2^B\}=\frac{i}{2}(\frac{\sigma_1\gamma_N+\sigma_N c_1\gamma_1}
{s_1\sigma_N\sigma_1}+i\frac{\gamma_1}{\sigma_1})D_1^A\de_{A,B}
\ee
Using this new constraint we find that the algebra closes linearly. If we
quantize the theory we should replace poisson-brackets by commutators.
Moreover, the constraint $D_1$ acts as the generator of time translations of
one unit of time. The constraint algebra of infinitesimal Lorentz
transformations and time steps of one unit now closes linearly.  This is
consistent with the fact that time should be quantized in this way in the
quantum theory as was noted by 't Hooft \cite{tHooft2,tHooft3}.\\
Finally we like to comment on the transitions that may occur during evolution
as one of the lengths $L_i$ shrinks to zero. Two out of nine possible
transitions are shown in figure (\ref{transitions}).
The rapidities $\eta_j$ of the newly opened edges can be calculated from the
rapidities of the pre-transition diagram. It is in fact these transitions that
render the theory non-trivial.\\
Next we will construct a toroidal spacetime using only one polygon.
In the next section we will give an alternative method to construct the torus.
We must cut the torus open according to figure (\ref{poltorus1}).
To do the cutting one could use the double line representation of
\cite{Franzosi}. Next we impose the constraint $C_1=0$:
\be
C_1=2(\al_1+\al_2+\al_3-2\pi)=0
\ee
This implies that both vertices $A$ and $B$ have no angle deficit:
$\al_1+\al_2+\al_3=2\pi$. Moreover, if we use this constraint in $C_2$ we see
that it is automatically fulfilled. It means that $C_2$ is no longer an
independent constraint. The only possible vertices that have no angle deficit
are vertices where all adjacent $\eta_j$ vanish or vertices where 2 rapidities
are equal and one vanishes (see figure (\ref{vertices})).

\noindent

Consider first the case where all rapidities are vanishing.
As the rapidities are zero, the $\al_i$ are free and we may choose them as
simple as possible: $\al_3=\pi$ and $\al_1=\al_2=\frac{1}{2}\pi$. (Note that
the free angles are not degrees of freedom.) The identification rules we have
found are thus (see figure (\ref{poltorus1}) and take all $\eta_j=0$):
\ba
x^\mu&=&[T_x(L_2)T_y(L_3)]^\mu_{~~\nu}x^\nu\\
x^\mu&=&[T_x(L_1+L_2)]^\mu_{~~\nu}x^\nu\nn
\ea
Comparison with (\ref{transstatic}) gives that this is precisely the static
torus as constructed using the methods of section 1.\\
Next consider the case where both vertices are of the type of figure
(\ref{vertices}). In this case only $\al_1=\al$ is free and we choose it to be
equal to $\frac{1}{2}\pi$ for convenience. Again we stress that it cannot be
considered as a degree of freedom as it ``separates'' two Lorentz frames that
are really the same ($\eta_3=0$). Different choices of $\al$ will therefore not
give different tori but just another way of describing the same
torus.\footnote{In this aspect we clearly disagree with the authors of
\cite{Franzosi}.} The torus that we constructed is thus an expanding or
shrinking torus in one direction and is therefore in the spacelike sector (see
figure (\ref{poltorus1}) and take $\eta_1=\eta_2=\eta$ and $\eta_3=0$).
Counting degrees of freedom we find $L_1,L_2,L_3,\eta$. As we have a closed
universe, time can not be defined at infinity, but must be defined in terms of
internal degrees of freedom. $L_3$ is the only degree of freedom that changes
(and it does so at a constant rate: $\frac{d}{dt}L_3=2\tanh\eta$), so we choose
it as our time variable:
\be
T=L_3
\ee
Another way of seeing this is that in generally covariant systems time
evolution is a gauge transformation connected with the freedom of
reparametrizing time. This implies that we should fix this gauge by choosing
explicitly time in terms of the phase space variables.\\
So we end up with an odd, three dimensional phase space which hints at the fact
that we missed part of the possible configurations by our choice of
slicing.\footnote{I would like to thank T. Jacobson for pointing this out to
me.} To see this more clearly we must analyze the identification rules and
compare them with (\ref{transdyn}). We find:
\ba
x^\mu&=&[T_x(L_1+L_2)]^\mu_{~~\nu}x^\nu\\
x^\mu&=&[T_x(L_2)T_y(L_3)B_y(\eta)]^{\mu}_{~~\nu}x^\nu\nn
\ea
Changing variables from $(L_1+L_2;L_2;\eta)$ to $(a_1;a_2;\eta_2)$ and
conjugating both transformations by:
\be
\Lambda_i\rightarrow T_t(\frac{L_3}{2v})T_y(\frac{-L_3}{2})\Lambda_i
T_t(\frac{-L_3}{2v})T_x(\frac{L_3}{2})~~~~~~~v=\tanh\eta
\label{tijd}
\ee
 we arrive at:
\ba
x^\mu&=&[T_x(a_1)]^\mu_{~~\nu}x^\nu\\
x^\mu&=&[T_x(a_2)B_y(\eta_2)]^\mu_{~~\nu}x^\nu\nn
\label{trafos}
\ea
The fact that we can transform $T_x(L_3)$ away is consistent with the fact that
it is simply time  which is not an independent degree of freedom. The
transformation (\ref{tijd}) is precisely a time translation to $t=0$ where the
identification is indeed given by (\ref{trafos}). It is now clear that we can
only access the phase space with $\eta_1=0$ using a one polygon slicing. The
next thing to try is then obviously to add another polygon in the construction
of the torus.

\section{The Two-Polygon Torus}

\begin{figure}[t]
\centerline{\psfig{figure=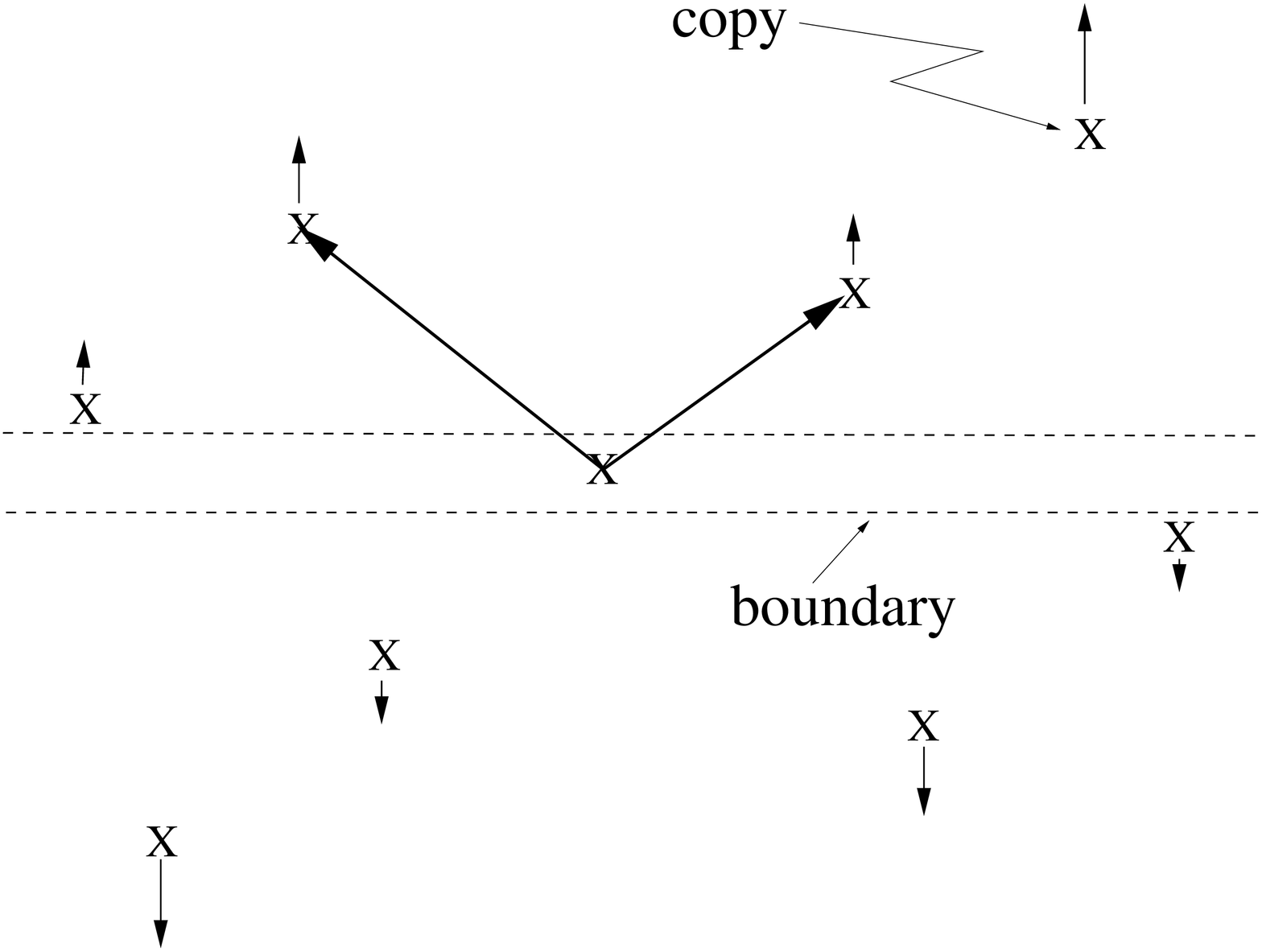,angle=-90,height=5cm}}
\caption{Lattice with moving copies for the torus.}
\label{lattice}
\end{figure}

In this section we discuss the general method for constructing a particle-free
torus. Using this
method due to 't Hooft \cite{tHooft2} we prove once more that one polygon is
certainly not enough to describe whole of phase space. We then discuss a two
polygon torus. In appendix B we show that its phase space is four dimensional.
\\
Let us define a Lorentz frame with an observer on it. The observer has
coordinates $x^\mu=(t,0,0)$. There are also Poincar\'e transformed copies of
the observer floating around. The Poincar\'e transformations of these copies
are given by combinations of the $\Lambda(\al_i)$ of (\ref{transdyn}) i.e.
$\Lambda(\al_1)^n\Lambda(\al_2)^m$ with $n$ and $m$ integers.\footnote{Remember
that $\Lambda(\al_1)$ en $\Lambda(\al_2)$ commute.} At $t=0$ all the copies are
situated at the x-axis with different velocities in the y-direction. As we let
the system evolve for a while the copies move up and are situated on a lattice
(see figure (\ref{lattice})).

\begin{figure}[t]
\centerline{\psfig{figure=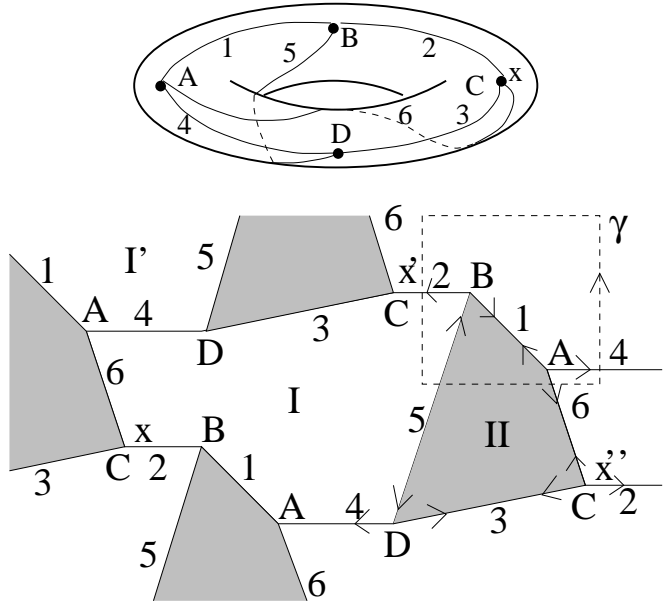,angle=-90,height=8cm}}
\caption{Torus made of two polygons.}
\label{2poltorus}
\end{figure}

\noindent
Using the method described in section 2 we can define a boundary between any
two copies in such a way that the coordinates of the copies are indeed
transformed according to the transformations $\Lambda_i$. If we choose to slice
spacetime with only one polygon we construct one boundary between each copy.
Where the boundaries meet, vertices are formed as described in section one. The
edges now enclose a region with precisely one copy on it. We may take one of
the regions as our fundamental region and identify points on the boundary. In
the case of the torus however all the boundaries are horizontal lines. Because
on the total lattice of copies one can always find a copy as close to the
y-axis as one pleases, the torus will become an infinitely small, infinitely
long cylinder. This is of course a singular situation and we will try to remedy
it by adding another polygon. In this case we may define two boundaries between
certain copies.
For the construction of such a two-polygon torus we add two extra vertex points
on the torus and cut it open along the lines drawn in figure (\ref{2poltorus}).
Of course there are many possibilities of drawing the lines between the
vertices corresponding with different kinds of two-polygon tori. Not all
cuttings are successful however in improving the singular situation of the
one-polygon torus. We also added  arrows on the boundaries. We will define a
boost to be positive as the arrows are directed towards the center (like in
$L_1$) and negative if they are directed towards the vertices (like in $L_5$).
The arrows are turned into vectors by giving them lengths equal to $|\eta_i|$.
For small $\eta_i$ the vectors should add up to 0 at a vertex due to the vertex
relations. For large $\eta_i$ this is not exactly true anymore but still a good
indication for how to draw the diagram approximately. The boundary between 2
polygons that represent the same piece of
(transformed) spacetime should be horizontal because observers see their copies
move in the $y$-direction (so $L_2$ and $L_4$ are horizontal). Moreover we
cannot change the values $\eta_2$ and $\eta_4$ by defining new Lorentz frames
since both frames (I and I') change by the same amount. In both figures we have
included an observer (X). Observer X sees his copy X' move in the $y$ direction
with velocity $v=\tanh(2\eta_2)$ and translated in the $x$-direction and
$y$-direction. As we have seen before, the translation in the $y$-direction is
due to the time-evolution and is immaterial. To see what the other independent
Poincar\'e transformation is we consider the copy X''. To see that this copy is
moving in the $y$-direction we consider the path $\gamma$. To get to the copy
we must move over the boundaries $L_5$ and $L_6$. But because the path $\gamma$
is contractable to a point (this is precisely what the vertex relations tell
us)  we see that the Lorentz transformation must be in the $y$-direction with
rapidity $2\eta_4-2\eta_2$. This proves that the observer X sees his copy X''
move in the $y$-direction with velocity $\tanh(2\eta_4-2\eta_2)$ and translated
in the $x$- and $y$-direction. The movement in the $y$-direction is once again
due to time evolution. If we would take for instance $\eta_2=\eta_4$ it is easy
to see that the translation in the $y$-direction vanishes. In the appendix we
give a detailed proof that the number of degrees of freedom is indeed four.
There are two configuration variables $L_2$ and $L_4$ and two independent
momentum variables $2\eta_2$ and $2\eta_4$. Using gauge transformations we can
choose the magnitude and orientation of $\eta_6$. Furthermore, the length $L_6$
will be taken as the physical time. Using this input we can uniquely calculate
the two-polygon torus of figure (\ref{2poltorus}). It is however hard to prove
that this particular two-polygon construction works for all possible tori
corresponding every Poincar\'e transformations of (\ref{transdyn}). In figure
(\ref{2poltorus}) we took the translations $a_i$ positive. For negative $a_i$
we might for instance consider the mirror-image of figure (\ref{2poltorus}). It
might even be necessary to add another polygon in certain regimes of phase
space. During time evolution 't Hooft-transitions may occur. This complicates
the description considerably.

\section{Discussion}
In this paper we studied the torus in the polygon representation. We found that
we need at least two polygons to describe the torus universe. This implies that
the description of \cite{Franzosi} covers only part of phase space. We also
gave an explicit construction of such a two-polygon representation. As soon as
particles are present, generic single polygon representations do exist.\\
Originally we had hoped that the torus was so simple that we could try to
quantize the system. It first seemed that we did not have to include
transitions in this model. As we try to quantize the two-polygon torus we can
follow two routes. The first is to try to find reduced phase space variables
and substitute commutators for poisson-brackets. However finding the effective
Hamiltonian will be a difficult job due to the non-linearity of the
constraints. Another possibility is to keep all classical variables and reduce
the Hilbert space by promoting the classical constraints to quantum operator
constraints. Of course, the non-linearity of the constraints will be hard to
handle also in this case. But even if one has overcome these problems one still
has to add the transitions as boundary conditions on the wave functions.
Moreover the torus should be invariant under the group of modular
transformations. There is a possibility however that we can simplify things a
bit. This was already suggested by 't Hooft in \cite{tHooft2}. Say we start
with a two-polygon torus. As the system evolves transitions will take place to
for instance to a three polygon representation. Using the gauge transformations
(generated by the constraints) we might be able to transform back to the
two-polygon representation. For instance, we could evolve one polygon forward
(or backward) in time until it disappears. Alternatively we could Lorentz
transform one polygon in such a way that it has the same Lorentz frame as a
neighboring polygon. The rapidity $\eta$ between the two-polygons disappears
and one is really left with one polygon less.\\ As all modular transformed
universes must be considered equal it might also happen that these modular
transformations can ``move us back'' to the original diagram. We did however
not investigate what the action of the modular group on the two-polygon torus
is.\\
A last issue concerning the quantization is the question if time is quantized.
For open systems where the coordinate $t$ can be taken as the physical time, 't
Hooft argued that time is indeed quantized \cite{tHooft2, tHooft3}. We found
that the constraint algebra supports this idea. For closed systems however, the
parameter $t$ cannot be considered as the physical time but merely as a gauge
parameter connected with reparametrization invariance. Whether the physical
time is also quantized in a closed system needs to be investigated further.
Waelbroeck seems to have found evidence that this is not the case
\cite{Waelbroeck2}. \\
To summerize we must conclude that the polygon approach is not the most
convenient way of treating the particle-free torus.

\section*{Acknowledgements}
I would like to thank G. 't Hooft and T. Jacobson for their many suggestions
and comments on this work. I also thank R. Drostery for interesting
discussions.
\appendix
\renewcommand{\thesection}{Appendix \Alph{section}:}
\section{Calculation of the Constraint Algebra}
\setcounter{equation}{0}
\renewcommand{\theequation}{\Alph{section}.\arabic{equation}}
\setcounter{figure}{0}
\renewcommand{\thefigure}{\Alph{section}.\arabic{figure}}
\begin{figure}[b]
\centerline{\psfig{figure=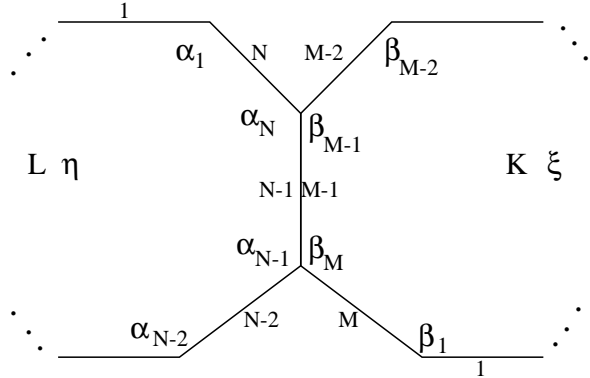,angle=-90,height=5cm}}
\caption{Definition of rapidities and angles for the calculation of the
constraint algebra.}
\label{definition}
\end{figure}
In this appendix we will give some details of the calculation of the algebra
(\ref{con-alg}). First we define the variables $L_i,\eta_i,\al_i$ for polygon I
and $K_j,\xi_j,\bt_j$ for polygon II in figure (\ref{definition}). The result
will not depend on the precise definition as one can check (i.e. where $L_1$ is
situated as compared to polygon II). Next we establish some basic results that
we will need in the following. Consider figure (\ref{vertex}). From the vertex
relations (\ref{vertexrelations}) we have the identity:\footnote{for the
definitions of $\sigma,\gamma,c,s$ see (\ref{def}).}
\be
c_i=\frac{\ga_i-\ga_j\ga_k}{\sg_j\sg_k}~~~~~~~~\mbox{and cyclic permutations}
\label{vertex1}
\ee
{}From this we calculate:
\ba
\frac{\dd\al_i}{\dd\eta_i}&=&-\frac{\sg_i}{s_i\sg_j\sg_k}\label{afgeleide1}\\
\frac{\dd\al_i}{\dd\eta_j}&=&-\frac{\sg_i c_k}{s_i\sg_j\sg_k}=
\frac{\sg_j\ga_k+c_i\sg_k\ga_j}{s_i\sg_j\sg_k}\label{afgeleide2}
\ea
If we take for instance the partial derivative $\frac{\dd}{\dd \eta_1}$ in the
above formula, we keep the other rapidities ($\eta_2$ and $\eta_3$) fixed.
Another useful identity that is derived using the last two identities is
(see figure (\ref{aap})):
\be
\frac{\dd\al_{k+1}}{\dd\eta_k}-i\frac{\ga_k}{\sg_k}=e^{-i\al_{k+1}}
(\frac{\dd\al_{k+1}}{\dd\eta_{k+1}}+i\frac{\ga_{k+1}}{\sg_{k+1}})
\label{identity}
\ee
A last (vertex) relation that we will need in the following is:
\be
s_i:s_j:s_k=\sg_i:\sg_j:\sg_k
\label{vertex2}
\ee
\begin{figure}[t]
\centerline{\psfig{figure=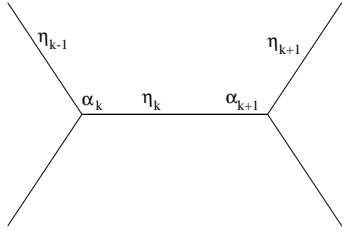,angle=-90,height=3cm}}
\caption{Definition of angles and rapidities for equality (\ref{identity}).}
\label{aap}
\end{figure}
The bracket $\{C_1^A,C_1^B\}$ must vanish because the angles only depend on the
momenta $\eta_i$. Next we calculate the following bracket between two different
polygons:
\ba
&&\{C_1^I,C_2^{II}\}=\\
&&\{\al_{N-1}+\al_N,e^{-i\vfi_{M-2}}(K_{M-2}-K_{M-1}e^{-i\bt_{M-1}}+
K_Me^{-i(bt_{M-1}+\bt_{M})})\label{br2}=\nn\\
&&\frac{1}{2}e^{-i\vfi_{M-2}}(\frac{\dd\al_N}{\dd\xi_{M-2}}-
\frac{\dd(\al_{N-1}+\al_N)}{\dd\xi_{M-1}}e^{-i\bt_{M-1}}+
\frac{\dd\al_{N-1}}{\dd\xi_M}e^{-i(\bt_{M-1}+\bt_M)})\nn
\ea
Note that only $\al_{N-1}$ and $\al_N$ contribute. Analogous to (\ref{C2}) we
used here the definition: $\vfi_i=\sum^i_{l=1}{\al_l}-(i-1)\pi$. Using
(\ref{afgeleide1}), (\ref{afgeleide2}) and (\ref{vertex2}) we find that it
vanishes. Next we calculate:
\ba
&&\{C_2^I,C_2^{II}\}=\\
&&e^{-i(\theta_{N-2}+\vfi_{M-2})}\{(L_{N-2}-L_{N-1}e^{-i\al_{N-1}}+
L_Ne^{-i(\al_{N-1}+\al_N)}),\nn\\
&&(K_{M-2}-K_{M-1}e^{-i\bt_{M-1}}+K_Me^{-i(\bt_{M-1}+\bt_M)}\}=\nn\\
&&-ie^{-i(\theta_{N-2}+\vfi_{M-2})}(L_Ne^{-i(\al_{N-1}+\al_N)}
\{\al_N,K_{M-2}\}+
L_{N-1}e^{-i(\bt_{M-1}+\al_{N-1})}\{\al_{N-1},K_{M-1}\}\nn\\
&& -e^{-i(\bt_{M-1}+\al_{N-1}+\al_N)}\{\al_{N-1}+\al_N,K_{M-1}\}
-L_{N-1}e^{-i(\bt_{M-1}+\bt_{M}+\al_{N-1})}\{\al_{N-1},K_M\}\nn\\
&&+L_Ne^{-i(\bt_{M-1}+\bt_{M}+\al_{N-1}+\al_N)}\{\al_{N-1},K_M\})~~~~
-~~~~(\bt\pijl\al~~,~~L\pijl K)\nn
\ea
All terms proportional to $L_i$ (or $K_i$) should vanish independently. The
calculation of the term proportional to $L_N$ (and $K_M$) turns out to be
proportional to the bracket (\ref{br2}) and therefore vanishes. For the
calculation of the remaining terms we notice that $L_{N-1}=K_{M-1}$. Using
again  (\ref{afgeleide1}), (\ref{afgeleide2}) and (\ref{vertex2}) we find that
it vanishes. The calculation of $\{C_2^I,\bar{C}_2^{II}\}$ follows the same
lines. Next we calculate the brackets between constraints of the same polygon:
\ba
&&\{C_1^I,C_2^I\}=\{\sum_{i=1}^{N}\al_i,\sum_{j=1}^N L_je^{-i\theta_j}\}\\
&&\frac{1}{2}((\frac{\dd\al_1}{\dd\eta_1}+\frac{\dd\al_2}{\dd\eta_1})-
e^{-i\al_2}(\frac{\dd\al_2}{\dd\eta_2}+\frac{\dd\al_3}{\dd\eta_2})+.....+
e^{-iC_1}(\frac{\dd\al_N}{\dd\eta_N}+\frac{\dd\al_1}{\dd\eta_N}))\nn
\ea
In the first term we add and subtract $i\frac{\ga_1}{\sg_1}$. Idem for the
second term, where we add and subtract $i\frac{\ga_2}{\sg_2}$ etc. Using
relation
(\ref{identity}) we see that pairs of terms cancel. The last term and the first
term must be calculated explicitly using (\ref{afgeleide2}) to find the desired
result (\ref{con-alg}). Finally we calculate:
\be
\{C_2^I,\bar{C}^I_2\}=\{\sum_{i=1}^N L_ie^{-i\theta_i},\sum_{j=1}^N L_j
e^{i\theta_j}\}
\ee
This can be simplified to:
\be
-i\sum_{i,j=1}^N L_j\cos(\theta_i-\theta_j)\frac{\dd\theta_j}{\dd\eta_i}
\ee
Again using (\ref{afgeleide2}) the desired result will follow.
\section{Degrees of Freedom for the \\  Two-Polygon Torus}
\setcounter{equation}{0}
\setcounter{figure}{0}
In this appendix we will carefully count the number of degrees of freedom
present in the two-polygon representation of the torus given in figure
(\ref{rechts}). First we notice that all the vertices and angles are already
present in and around polygon II (see figure (\ref{rechts})).\footnote{ The
notation for the angles is defined in figure (\ref{rechts})}
If we count the momentum degrees of freedom we only have to consider this part
of the diagram. Once all rapidities and angles are determined here, they are
also fixed in the rest of the diagram. We claim that the momentum degrees of
freedom are $\eta_2$ and $\eta_4$. They can not be changed by gauge
transformations as explained in section 3. Different choices of $\eta_2$ and
$\eta_4$ imply different tori. By using Lorentz transformations we can choose
the magnitude and orientation of $\eta_6$. This determines the relative Lorentz
frames of I and II. So we have now $\eta_2$, $\eta_4$, $\eta_6$, $A_1$ and
$C_3$. (Remember that $L_2$ and $L_4$ must be horizontal.) Given two $\eta_i$
and one angle around a vertex we can determine all angles and rapidities at
that vertex using the vertex relations. It implies that we can calculate now
$A_6$, $A_4$, $\eta_1$, $C_3$, $C_6$ and $\eta_3$. Using the constraint that
$A_6+B_5=2\pi$ we can determine $B_5$. But now we have two rapidities ($\eta_2$
and $\eta_1$) and one angle ($B_5$) at the vertex B. So we can calculate $B_1$,
$B_2$ and $\eta_5$. Because the angles inside polygon II must add up to $2\pi$
we know $D_4$. Again we have two rapidities and one angle at vertex D which
determines $\xi$, $D_5$ and $D_3$. Of course we would like to prove now that
$\xi=\eta_4$ and its orientation to be horizontal. Then the constraints
$B_1+D_3=\pi$ and $D_5+C_6=2\pi$ are also obeyed. We calculate $\xi$ and its
orientation by noting that the holonomy around the vertex D must be trivial
(see \ref{vertexrel}). From that we derive the vertex relations. As long as the
vertex relations are obeyed one can move the loop of the holonomy over a vertex
without changing its holonomy. Let's change the loop around vertex D to the
loop $\ga '$. We find for the holonomy:
\be
B(2\eta_5)B(2\eta_6)B(-2\eta_2)B(2\xi)=I
\ee
Consider also the holonomy around loop $\ga$ if we traverse it in the direction
as is indicated in the figure:
\be
B(2\eta_5)B(2\eta_6)B(2\eta_4)B(-2\eta_4)=I
\ee
Because $B(2\eta_2)$ and $B(2\eta_4)$ commute we derive that
$B(2\xi)=B(2\eta_4)$ which proves the statement. As mentioned earlier we can
finish the rest of the diagram (except for the fact that we have to choose the
lengths $L_2$ and $L_4$) because all angles are known now. We are also assured
that the angles inside polygon I will add up to $6\pi$. To see this we notice
that the constraint splits actually in four constraints:
\ba
&&C_3+A_1=\pi\\
&&D_5+C_6=2\pi\nn\\
&&B_1+D_3=\pi\nn\\
&&B_5+A_6=2\pi\nn
\ea
But these constraints were used or verified in the previous proof so they are
automatically obeyed.
\begin{figure}[t]
\centerline{\psfig{figure=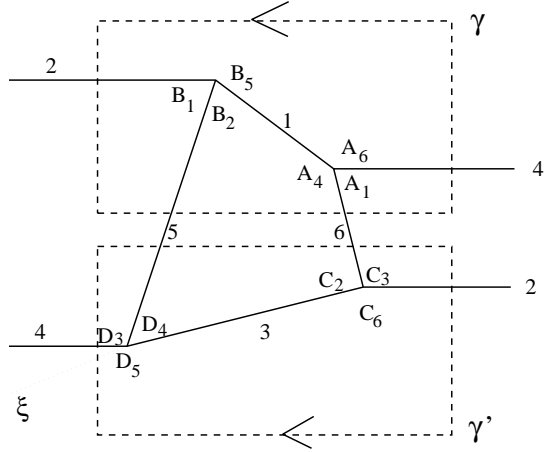,angle=-90,height=6cm}}
\caption{Part of the two-polygon torus.}
\label{rechts}
\end{figure}

\noindent
Next we count the independent length variables. It is clear that given a
diagram we can change the lengths $L_2$ and $L_4$ independently and still have
a valid diagram. So to make life easy we choose them to be zero. We have now
two polygons with four of the same sides but in general different internal
angles.
Given the size of one of these sides (say $L_6$) and the internal angles there
is  generally only one solution for the remaining three edge lengths for which
we can draw these two ``four-gons''.  The only independent variable (the scale
or $L_6$) will be used to define the physical time. This concludes the proof
that $L_2$ and $L_4$ are the only two degrees of freedom in configuration
space.


\begin{thebibliography}{99}
\bibitem{eerst} Staruszkiewicz A 1963 {\em Acta Phys. Polon.} {\bf 24} 734
\bibitem{beginarticle} Deser S, Jackiw R and 't Hooft G 1984 {\em Ann. Phys.}
{\bf 152} 220
\bibitem{tHooft1}  't Hooft G 1992 {\em Class. Quant. Grav.} {\bf 9}  1335\\
                   't Hooft G 1993 {\em Class. Quant. Grav.} {\bf 10} S79 \\
                   't Hooft G 1993 {\em Class. Quant. Grav.} {\bf 10} 1023
\bibitem{tHooft2}  't Hooft G 1993 {\em Class. Quant. Grav.} {\bf 10} 1653
\bibitem{tHooft3}  't Hooft G 1996 {\em Quantization of Point Particles in 2+1
Dimensional
                    Gravity and space-time discreteness.} {\bf gr-qc/9601014}
\bibitem{Witten} Achucarro A and Townsend P 1986 {\em Phys. Lett.}
                  B{\bf 180} 85\\
                 Witten E 1988 {\em Nucl. Phys.} B{\bf 331} 46
\bibitem{Welling} Welling M 1996 {\em Class. Quant. Grav.} {\bf 13} 653\\
                  Welling M and Bijlsma M 1996 {\em Pauli-Lubanski Scalar in
the Polygon
                  Approach to (2+1)-Dimensional Gravity.}  {\bf
gr-qc/9601025}\\
                  Welling M 1995 {\em Some Approaches to (2+1)-Dimensional
Gravity coupled
                                   to Point Particles.} {\bf hep-th/9511211}
\bibitem{Carlip1} Carlip S 1990 {\em Phys. Rev.} D{\bf 42} 2647
\bibitem{Carlip2} Carlip S 1993 {\em Phys. Rev.} D{\bf 47} 4520\\
                  Carlip S 1992 {\em Phys. Rev.} D{\bf 45} 3584\\
                  Carlip S 1991 {\em Class. Quant. Grav.} {\bf 8} 5
\bibitem{Louko}   Louko J and Marolf D M 1994 {\em Quant. Class. Grav.} {\bf
11} 311
\bibitem{Moncrief} Moncrief V 1989 {\em J. Math. Phys.} {\bf 30} 2907
\bibitem{Nakao}  Hosoya A and Nakao K 1990 {\em Class. Quant. Grav.} {\bf 7} 63
\bibitem{Waelbroeck1}  Criscuola A and Waelbroeck H 1996 {\em A Hamiltonian
Lattice Theory
                       for homogeneous curved Space-Times in (2+1)-Dimensions.}
                       {\bf gr-qc/9601028}\\
                       Criscuola A and Waelbroeck H 1995 {\em Quantization of
(2+1) Gravity
                       on the Torus.} {\bf gr-qc/9509041}
\bibitem{Waelbroeck2}  Waelbroeck H and Zapata J A 1996 {\em (2+1) Covariant
Lattice Theory
                       and 't Hooft's Formulation.} {\bf gr-qc/9601011}
\bibitem{Franzosi} Franzosi R and Guadagnini E 1996 {\em Class. Quant. Grav.}
{\bf 13} 433
\bibitem{Nelson}   Nelson J E and Regge T 1991 {\em Phys. Lett.} B{\bf 272}
213\\
                   Nelson J E and Regge T 1989 {\em Nucl. Phys.} B{\bf 328} 190
\bibitem{Ezawa}    Ezawa K 1994 {\em Phys. Rev.} D{\bf 50} 2935\\
                   Ezawa K 1994 {\em Phys. Rev.} D{\bf 49} 5211\\
                   Ezawa K 1993 {\em Reduced Phase Space of the First Order
Einstein Gravity
                   on $R\times T^2$} {\bf hep-th/9312151}
\end{thebibliography}
\end{document}